\documentclass[sigplan,nonacm]{acmart}

\setcopyright{none}
\acmConference{}{}{}
\acmBooktitle{}
\acmDOI{}
\acmISBN{}

\usepackage{booktabs}
\usepackage{tabularx}
\usepackage{array}
\usepackage{xspace}
\usepackage{graphicx}
\usepackage{subcaption}
\usepackage{tabularray}
\usepackage{algorithm}
\usepackage{algpseudocode}
\usepackage{enumitem}
\usepackage{amsmath}
\usepackage{xurl}

\AtBeginDocument{%
  }

\newcommand{\system}{\textsc{Tessera}\xspace}
\newcommand{\appref}[1]{Appendix~#1}

\newcolumntype{Y}{>{\raggedright\arraybackslash}X}

\newif\ifincludeevaluation
\includeevaluationtrue

\usepackage{xcolor}

\makeatletter
\patchcmd{\@typeset@author@bx}
  {\@authorfont\@currentauthors\par\@affiliationfont}
  {\@authorfont{\def\par{\def\par{\quad}}\@currentauthors}\par\@affiliationfont}
  {}{\PackageError{tessera}{Could not configure the shared author line}{}}
\makeatother

\hypersetup{keeppdfinfo=true}

\begin{document}

\title{Decoupling Logical Masks from GPU Execution for Dynamic Block-Sparse Attention}

\author{Shanghao Liu}
\author{Xiaoyun Yu}
\author{Wanting Li}
\author{Wenqi Jiang\textsuperscript{*}}
\affiliation[obeypunctuation=true]{%
  \institution{National University of Singapore}\\
  \city{Singapore}, \country{Singapore}\\
  \href{mailto:wenqi.jiang@nus.edu.sg}{\texttt{wenqi.jiang@nus.edu.sg}}\textsuperscript{*}
}
\renewcommand{\shortauthors}{Liu et al.}

\sloppy
\begin{abstract}
  Attention computation makes inference expensive in \emph{video diffusion transformers} (vDiTs), which generate videos through iterative denoising.
\emph{Block-sparse attention} (BSA) reduces this cost by computing only blocks selected by a \emph{logical mask}, which specifies the attention interactions to compute.
Existing BSA runtimes either fix one physical execution plan per block geometry, which limits adaptation to varying masks and graphics processing units (GPUs), or plan each mask at runtime, which adds preparation overhead to every request. Because masks change across requests, this per-mask planning must be repeated, so its cost can outweigh the execution-time savings.
We present \system{}, a runtime for dynamic BSA that decouples logical masks from GPU execution through two-layer physical execution plans that preserve the specified attention interactions.
First, its physical mapping layer retains, combines, or subdivides logical blocks into physical tiles suited to different mask shapes and GPU architectures.
Second, its task organization layer groups and schedules tiles within GPU tasks to reuse data, expose parallelism, and overlap data movement with computation.
Third, its plan selector chooses a plan for each request with low overhead, through either a plan table built from offline profiling or a calibrated cost model, which extends selection to masks outside the profiled distribution and to a new GPU with about 100 calibration requests.
Implemented with specialized CUDA kernels for four NVIDIA GPU generations and evaluated on 2,312 real attention masks and two video diffusion models, \system{} achieves up to $2.88\times$ BSA request speedup over baseline systems in these models and speeds up their 50-step loop by $1.21$--$1.24\times$.

\end{abstract}

\maketitle

\section{Introduction}
\label{sec:intro}

Video generation powered by video diffusion transformers (vDiTs) has emerged as a popular yet computationally demanding generative AI application~\cite{deepmind_veo31,seedance2026seedance2}.
For example, generating a five-second 720p clip with HunyuanVideo on a single H100 GPU takes 945 seconds~\cite{zhang2025sta}, equivalent to 3.15 GPU-minutes per second of generated video, and attention accounts for roughly 85\% of this runtime~\cite{zhang2025sta}.
At each denoising step, the attention layers of a vDiT process the token sequence of the full spatiotemporal latent---115K tokens for a five-second clip~\cite{zhang2025sta}---and the attention cost grows quadratically with sequence length~\cite{xi_sparse_2025}.

\emph{Block-sparse attention} (BSA) is a widely adopted approach to reduce this cost, used by methods such as VSA, SpargeAttn, SVG, SVG2, and SPADE~\cite{zhang_vsa_2025,zhang_spargeattn_2025,xi_sparse_2025,yang_sparsevideogen2_2025,liu2026spade} and by industrial video models such as HunyuanVideo-1.5 and LongCat-Video~\cite{wu2025hunyuanvideo15,longcat2025technical}.
BSA divides the attention matrix into rectangular \emph{logical blocks}, each covering $B_Q$ query tokens and $B_{KV}$ key/value tokens, with block geometry $B=(B_Q,B_{KV})$.
A \emph{logical mask} $M$ marks which logical blocks are active, and BSA computes only those blocks.

The highly dynamic masks used in BSA make efficient GPU execution challenging.
Active blocks can change across video-generation requests, attention heads, layers, and denoising steps, while logical block sizes are parameters chosen by the sparsity algorithm~\cite{wu2025hunyuanvideo15,longcat2025technical,zhang_vsa_2025,liu2026spade,yang_sparsevideogen2_2025}.
The active blocks per query row and the block geometry set the work on inactive entries, the metadata traffic, and the load balance of a kernel, yet a GPU kernel fixes its tiling and schedule at compile time, before the mask is known.
Each BSA invocation, termed a \emph{runtime request}, therefore requires an execution plan suited to its mask structure, block geometry, input tensors, and target GPU, and every request pays the cost of choosing and preparing that plan.
A BSA runtime must thus choose, for every request, a plan that fits its mask and GPU, at a cost well below the attention time that the plan saves.

\textbf{Existing approaches tie the physical plan to logical blocks, which misses faster executions, or reuse it only for repeated inputs, which keeps planning on the request path.}
Specialized kernels such as Block Sparse Attention and flex-block-attn~\cite{guo2024blocksparse,peng2025flexblockattn} are precompiled and keep request overhead low, but they fix one execution strategy for preset block sizes, whereas the fastest physical execution changes with the mask: within the H100 masks of one block geometry, $B=(32,16)$, the fastest plan uses larger tiles for 197 masks and tiles equal to the logical blocks for 53 (Section~\ref{sec:eval-adapt}).
FlexAttention and FlashInfer~\cite{dong2024flexattention,ye2025flashinfer} compile kernels per attention variant and adapt to each mask at runtime, but their tiles stay tied to logical blocks, which limits the executions they can reach (Section~\ref{sec:semantics-vs-execution}).
FlashInfer also computes its scheduling plan on the CPU for every new mask and reuses it only across operators that share its input layout~\cite{ye2025flashinfer}.
Sparse tensor compilers generate code for a chosen format and schedule~\cite{kjolstad2017taco,ye2023sparsetir,won2025insum}, and input-aware autotuners choose the format and schedule that fit each sparse input~\cite{vuduc2005oski,won2023waco}; like FlashInfer's plan, this specialization is amortized over repeated runs on the same input.
Dynamic BSA breaks this premise: in the Wan2.1 and HunyuanVideo-1.5 generations of Section~\ref{sec:eval-boundary}, none of the 11,402 and 2,700 runtime requests repeats the mask of an earlier request.
Per-input planning and search therefore stay on the request path: on H100, searching the plans online for each mask makes cold requests much slower than a fixed kernel with one tile per logical block (Section~\ref{sec:eval-selection}).
These compilers and autotuners also target additive reductions, such as sparse matrix--dense matrix multiplication (SpMM) and sampled dense--dense matrix multiplication (SDDMM), whereas attention reduces via an online softmax, whose state flow changes when tiles are remapped (Section~\ref{sec:block-to-tile}).

\textbf{We present \system{}, a runtime that turns the physical plan of dynamic block-sparse attention into a per-request decision and moves the plan search offline.}
\system{} separates the algorithm's logical attention specification from its GPU implementation through a semantics-preserving, two-layer \emph{physical execution plan} (physical plan) and a \emph{plan selector}.
First, the \emph{physical mapping} layer decouples logical blocks from physical tiles, allowing the same mask to be executed using tiles that retain, combine, or subdivide logical blocks while preserving the specified attention interactions.
Second, the \emph{task organization} layer determines how these tiles are grouped into GPU tasks, each executed by one GPU thread block, how data movement overlaps with computation, and how partial results are reduced.
Together, the two layers define the physical plan that executes the mask.
Third, the \emph{plan selector} ranks the eligible plans of a precompiled native CUDA catalog for each request by request features and mask statistics rather than by mask identity, and returns the highest-ranked eligible plan, so offline profiling serves masks that never recur.
It offers two methods: a \emph{plan table}, the default, which uses offline profiling to rank plans per \emph{performance regime}, a group of requests that shares a profiled plan ranking, and a \emph{calibrated cost model}, which predicts each plan's latency from the mask's counts.

\begin{figure}[t]
  \centering
  \includegraphics[width=\columnwidth,trim=0 29bp 0 45bp,clip]{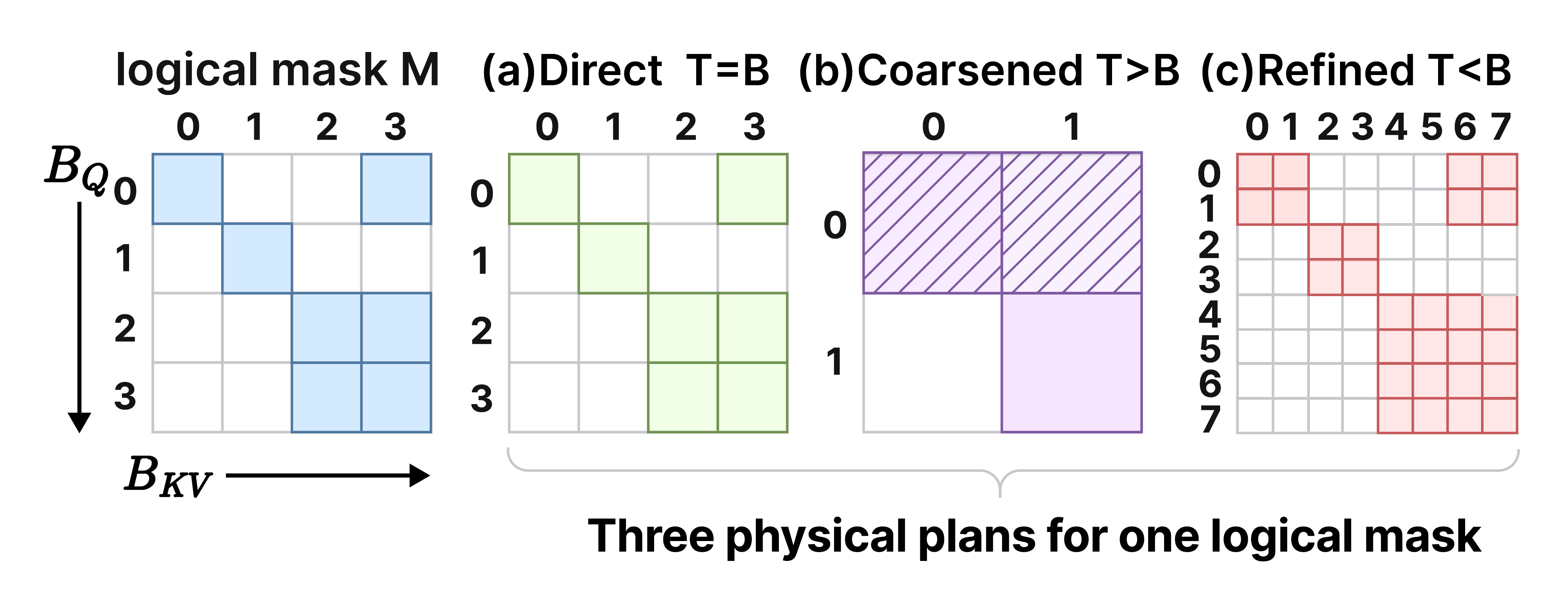}
  \caption{The same logical sparse-attention mask (colored blocks in the left sub-figure) can be executed using physical tiles that are (a) equal to, (b) larger than, or (c) smaller than the logical blocks.
    Here, $B$ denotes the logical block size, and $T$ denotes the physical tile size used for GPU execution.}

  \Description{A schematic showing one block-sparse mask on the left and three
    physical plans on the right that cover the same active region with equal,
    larger, and smaller tiles. Each plan is drawn on its own tile grid, whose
    axis labels count physical tiles rather than logical blocks.}
  \label{fig:intro-plans}
\end{figure}

\textbf{First, flexible mapping from logical blocks to physical tiles adapts tile geometry to the mask and the GPU.}
Let $T=(T_Q,T_{KV})$ denote the physical tile geometry: each tile covers the interactions between $T_Q$ query tokens and $T_{KV}$ key/value tokens that the GPU processes at a time.
\emph{Direct} keeps $T=B$ (Figure~\ref{fig:intro-plans}(a)), \emph{Coarsened} covers several logical regions with a larger tile and masks out inactive entries (Figure~\ref{fig:intro-plans}(b)), and \emph{Refined} divides logical blocks into smaller tiles (Figure~\ref{fig:intro-plans}(c)).
Coarsening improves operand reuse at the cost of computation on masked-out entries, whereas refinement reduces the per-tile footprint at the cost of more data movement and tiles.
Hence the best mapping depends on the mask and GPU (Sections~\ref{sec:why-ranking-changes} and~\ref{sec:eval-adapt}).

\textbf{Second, the task organization layer groups physical tiles into GPU tasks and orders their execution.}
Grouping tiles into one task reuses the query tile and output accumulator but leaves fewer tasks for the streaming multiprocessors (SMs). Conversely, splitting a query tile's K/V reduction across tasks exposes parallelism but adds a merge, and deeper pipelines increase overlap but need more buffering.
At a fixed mapping on H100, task organization further reduces kernel latency (Section~\ref{sec:eval-adapt}).

\textbf{Third, the plan selector chooses a high-performance physical plan with low overhead, using either of two methods.}
Exhaustively preparing and probing eligible plans for each new mask is expensive, as noted above.
\system{} moves this search offline.
With the plan table, it profiles the target GPU's plans on a representative mask corpus and ranks them per performance regime, so a ranking also serves masks never seen during profiling.
On H100, the plans that \system{} selects match the kernel speedup of exhaustive online search, and \system{} serves cold requests $14.42\times$ faster (Section~\ref{sec:eval-selection}).
The calibrated cost model fits a few parameters per plan and GPU to measured latencies, extending selection to masks outside the profiled distribution and to a new GPU with about 100 calibration requests (Section~\ref{sec:cost-model}).

We evaluate \system{} across four NVIDIA GPU generations and 2,312 real attention masks.
On every GPU platform, \system{} reduces the geometric-mean complete-request latency across workload families relative to every baseline system.
Across these attention masks, \system{} achieves geometric-mean speedups of $1.57$--$1.76\times$ over FlashInfer, $2.01$--$3.16\times$ over FlexAttention, and $2.00\times$ over the flex-block-attn library across the four workload families that flex-block-attn supports on H100.
\system{} also achieves kernel-only geometric-mean speedups of $1.16$--$4.91\times$ over the baselines across all four platforms.
We further evaluate \system{} within 720p Wan2.1 and HunyuanVideo-1.5 inference on H100, where it accelerates attention by $2.17$--$2.88\times$ and the full 50-step diffusion loop by $1.21$--$1.24\times$.

This paper makes three contributions:
\begin{itemize}[leftmargin=1.2em]
  \item \textbf{Two-layer physical plans that decouple logical masks from execution.}
        Direct, Coarsened, and Refined mappings retain, combine, or subdivide logical blocks while preserving the masked-attention semantics under the online softmax, and task organization schedules the tiles; native CUDA catalogs cover four GPU generations.

  \item \textbf{A plan selector independent of mask identity, with two methods.}
        A plan table amortizes offline profiling across all masks that share a performance regime, and a calibrated cost model generalizes to masks outside the profiled distribution and ports selection to a new GPU with about 100 calibration requests.

  \item \textbf{The \system{} runtime and its evaluation.}
        We evaluate \system{} on four GPUs, 2,312 real attention masks, and end to end in two industrial video diffusion models.
\end{itemize}

\section{Background and Motivation}
\label{sec:problem}

\subsection{Dynamic Masks and Logical Block Geometry}
\label{sec:dynamic-bsa}

The logical mask specifies the attention interactions that each execution must preserve.
For one attention head, let $Q\in\mathbb{R}^{N_q\times d}$ be the query
matrix and $K,V\in\mathbb{R}^{N_{kv}\times d}$ the key and value matrices,
where $N_q$ and $N_{kv}$ are sequence lengths and $d$ is the head dimension.
A block-sparse mask $M$ marks a set of logical block pairs active, each
covering $B_Q$ query tokens and $B_{KV}$ key/value tokens under the logical
block geometry $B=(B_Q,B_{KV})$.  Expanding the active pairs to token
coordinates gives the \emph{effective interaction set} $E(M)$, and masked
attention computes
\begin{equation}
  O_M=
  \operatorname{softmax}\!\left(
  \frac{QK^{\mathsf T}}{\sqrt d}+\mathcal{B}_M
  \right)V,
  \label{eq:masked-attention}
\end{equation}
where $\mathcal{B}_M(i,j)=0$ for $(i,j)\in E(M)$ and $-\infty$ otherwise.
$E(M)$ specifies the required interactions independently of how the work is
laid out on the GPU.

The \emph{sparsifier}, which selects the active attention blocks, supplies
both $M$ and $B$ to the runtime.  Video sparsifiers recompute the active
block pairs across runtime requests, heads, layers, and denoising steps,
and some fix $B$ while others select it per
model~\cite{zhang_spargeattn_2025,xi_sparse_2025,zhang_vsa_2025,
  wu2025hunyuanvideo15,longcat2025technical,liu2026spade,
  yang_sparsevideogen2_2025}.

\subsection{The Cost of Logical-to-Physical Coupling}
\label{sec:semantics-vs-execution}

The logical block is the sparsifier's choice of where attention is
computed, whereas the physical tile is the kernel's choice of how that work
is laid out on the GPU; a backend can nevertheless couple the two.  The
\emph{physical tile geometry} $T=(T_Q,T_{KV})$ defines the
query--key/value region processed at a time, the native instruction
geometry defines the shape consumed by a Tensor Core instruction, and a
\emph{GPU task} groups one or more physical tiles for scheduling.

Existing execution paths retain constraints from logical block geometry.
FlexAttention's standard Triton forward path requires $B_Q$ and $B_{KV}$ to
be divisible by $T_Q$ and $T_{KV}$, so tiles can match or subdivide
logical blocks~\cite{dong2024flexattention}, and FlashInfer's block-sparse
design aligns block rows with query tiles~\cite{ye2025flashinfer}.
Direct mapping, $T=B$, simplifies sparse execution
because an active block index identifies a physical tile without metadata
describing membership inside that tile.  However, the sparsifier selects
$B$ for mask expressiveness rather than for the GPU, so the same $E(M)$
often executes slower under $T=B$ than under an alternative physical
covering (Section~\ref{sec:eval-adapt}).
\system{} exposes mappings that retain, combine, or subdivide logical blocks
while preserving $E(M)$ (Figure~\ref{fig:intro-plans}), and selects among
eligible precompiled plans for each runtime request with its plan selector
(Section~\ref{sec:regime-selection}).

\subsection{Data--Compute Pipeline Cost}
\label{sec:pipeline-view}

As in FlashAttention-3~\cite{flashattention3} and
CUTLASS~\cite{nvidia_cutlass_gemm}, block-sparse attention can execute as a
two-stage \emph{data--compute pipeline} (\appref{A}).  The
\emph{data-movement stage} transfers operands and mask metadata between
global memory and on-chip storage, and the \emph{attention-compute stage}
computes $QK^\top$, updates the softmax, and multiplies the attention
weights by $V$.  With overlapped stages, the longer stage sets
steady-state throughput.
For architecture $A$, mask $M$, and plan $P$,
\begin{equation}
  \begin{split}
  t_A(P,M)\approx{}&
  \max\!\left(
  \frac{D_P(M)}{\beta^{\mathrm{eff}}_A(P,M)},\;
  \frac{F_P(M)}{\pi^{\mathrm{eff}}_A(P,M)}
  \right)\\
  &+t_{\mathrm{aux}}(P,M).
  \end{split}
  \label{eq:pipeline-model}
\end{equation}
Equation~\ref{eq:pipeline-model} separates stage demand from the GPU's
service rate.  $D_P$ and $F_P$ count the bytes and arithmetic of
the main attention kernel: $D_P$ includes Q/K/V transfers, the final-output
store, metadata, and plan-specific state, and $F_P$ includes work on
inactive entries covered by physical tiles.  The effective bandwidth
$\beta^{\mathrm{eff}}_A$ and compute rate $\pi^{\mathrm{eff}}_A$ depend on
$P$ and $A$~\cite{williams2009roofline}.  The term $t_{\mathrm{aux}}$ covers
overheads outside the steady state, such as scheduling, pipeline
fill and drain, and any separate reduction.
Physical mapping and task
organization change both demands and effective rates, and thus
the cost of executing the same $E(M)$.

\subsection{Why the Fastest Mapping Changes Across GPUs}
\label{sec:why-ranking-changes}

The fastest plan for a mask depends on the GPU, and the dependence is large
enough to matter when plans are shared across GPUs.
Table~\ref{tab:mapping-flip} isolates the GPU dependence.

\begin{table}[t]
  \centering
  \caption{Validation GM regret when the target GPU runs the source GPU's fastest mapping type.}
  \label{tab:mapping-flip}
  \small
  \setlength{\tabcolsep}{2pt}
  \begin{tabular*}{\columnwidth}{@{\extracolsep{\fill}}lccc@{}}
    \toprule
    Source $\rightarrow$ target & A100 & RTX~4090 & H100 \\
    \midrule
    A100 & -- & 1.023 & 1.215 \\
    RTX~4090 & 1.008 & -- & 1.082 \\
    H100 & 1.097 & 1.062 & -- \\
    \bottomrule
  \end{tabular*}
\end{table}

\textbf{Cross-GPU transplant.}  Table~\ref{tab:mapping-flip}
uses the profiling requests held out from table fitting (validation
requests) on A100, RTX~4090, and H100.  For each request, the target GPU
runs its own fastest plan within the mapping type that is fastest on the source GPU, and the table reports the
geometric-mean (GM) regret, its latency relative to the target GPU's own
fastest plan.  Every transplant
is slower; the GM regret reaches $1.215$ from A100 to H100.

\textbf{Implementation.}  The fastest plan on a GPU also depends on how its
native implementation uses architecture-specific features: on H100, a
warp-specialized Direct kernel beats Refined on 44 of 56 non-tied masks with
the SM90 instruction \texttt{setmaxnreg} and on none without it
(\appref{B}).  A ranking measured on one GPU therefore does not carry over
to another.
A per-GPU lookup would recover most of this loss: for transplants from A100
and RTX~4090 to H100, 97.1\% and 98.8\% of the summed transplant
\emph{log-regret} (log of regret, Section~\ref{sec:eval-setup}) falls on requests whose fastest mapping type
on H100 is the most frequent one on H100 for the same logical geometry.
Section~\ref{sec:regime-selection} therefore builds the plan selector per
GPU, by profiling or calibration on that GPU.

\section{\system{} Overview}
\label{sec:overview}

\begin{figure*}[t]
  \centering
  \includegraphics[width=0.92\textwidth]{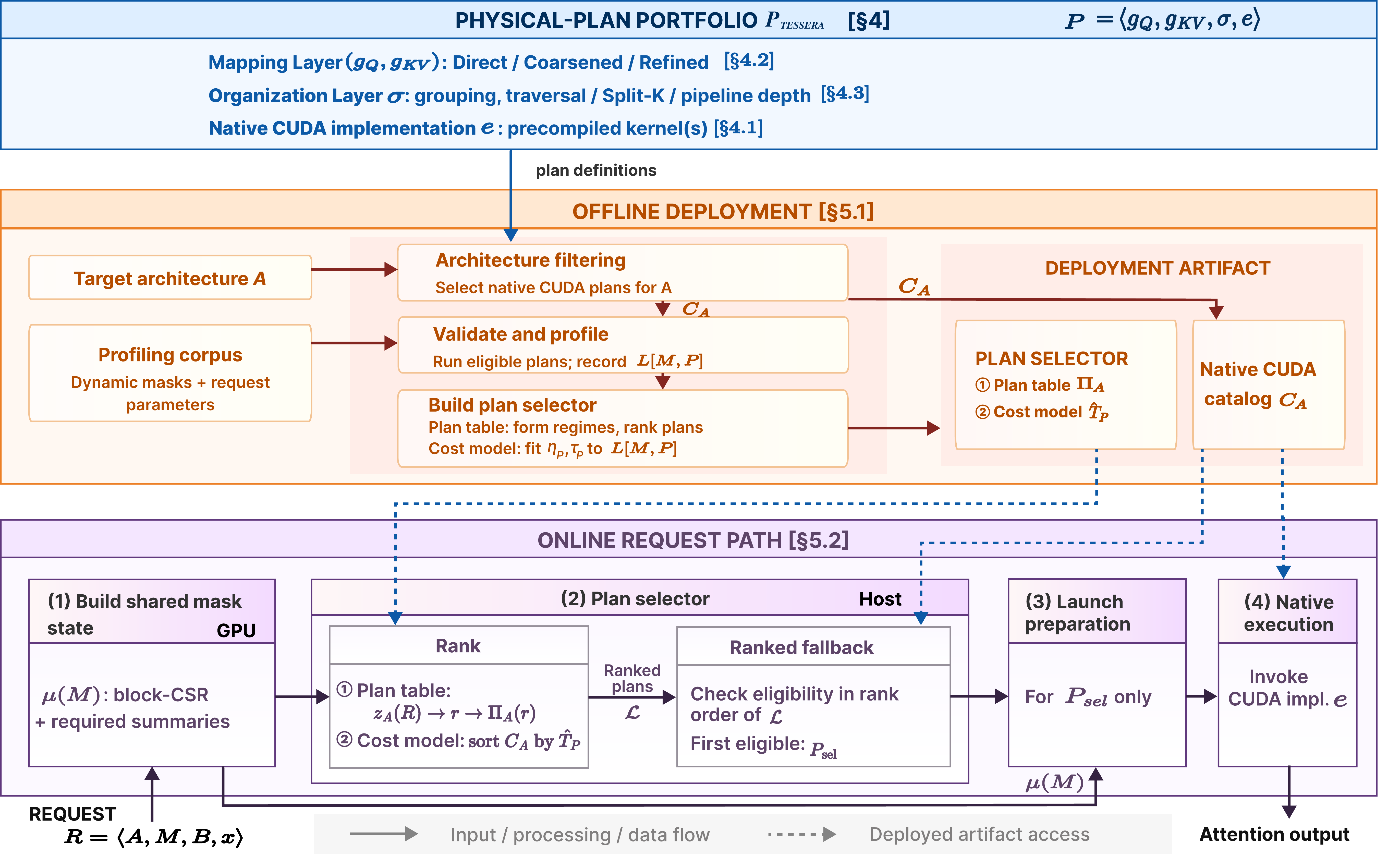}
  \caption{\system{} overview.  Offline deployment filters the
    physical-plan portfolio into the native CUDA catalog $C_A$ of the
    target architecture $A$, then validates and profiles the catalog plans
    on a profiling corpus and builds the plan selector with one of two
    methods: the plan table $\Pi_A$ (default) or a calibrated cost model.
    Online, the runtime builds the shared mask state $\mu(M)$ on the GPU;
    the plan selector ranks the plans and selects the first eligible one;
    the runtime prepares only that plan and invokes its native CUDA
    implementation.}
  \Description{A system diagram with three horizontal bands.  The top
    band, the physical-plan portfolio, lists each plan as a mapping layer,
    an organization layer, and a precompiled native CUDA implementation.
    The middle band, offline deployment, takes the target architecture and
    the profiling corpus, filters the portfolio into the native CUDA
    catalog through architecture filtering, validates and profiles the
    eligible plans, builds the plan selector (the plan table or a
    calibrated cost model), and emits the catalog and plan selector as one
    versioned deployment artifact.  The bottom band, the
    online request path, takes a request through building the shared mask
    state on the GPU, plan selection on the host with ranking (regime
    lookup in the plan table or prediction by the cost model) and ranked fallback, launch preparation of the selected plan, and native
    execution that produces the attention output.}
  \label{fig:system-overview}
\end{figure*}

\system{} separates the construction of physical plans for dynamic
block-sparse attention from their selection at runtime
(Figure~\ref{fig:system-overview}): the plan selector evaluates the plans
offline, by profiling or by calibration, and ranks them online without
probing candidate plans on the request path, so \system{} adapts execution
to each mask and GPU at low cost.

\textbf{Physical-plan portfolio.}
The \emph{physical-plan portfolio} defines the available execution choices
(Figure~\ref{fig:system-overview}, top).  Each choice is a precompiled CUDA implementation of a
mapping layer and an organization layer.  The mapping layer determines which
physical tiles cover the active logical blocks.  The organization layer
groups those tiles into GPU tasks and sets their traversal and parallel
execution.

\textbf{Offline deployment.}
Offline deployment converts the portfolio into architecture-specific plans
and rankings (Figure~\ref{fig:system-overview}, middle).  For a target architecture $A$, \system{}
first filters the portfolio for plans supporting $A$, producing the
\emph{native CUDA catalog} $C_A$.  It then validates and profiles these
plans using dynamic masks and request parameters from video models.  From these measurements, \system{} builds the plan
selector with one of two methods (Section~\ref{sec:policy-construction}).  The plan table $\Pi_A$, the default, groups requests
with similar plan performance into performance regimes and stores each regime's ranking.  The calibrated cost model fits
per-plan parameters to the measured latencies.  The catalog and the plan selector are versioned and deployed together.

\textbf{Online request path.}
The online request path uses the deployed catalog and plan selector to select and
execute one plan for the current request (Figure~\ref{fig:system-overview}, bottom).  A request $R$ provides the logical
mask $M$, the logical block geometry $B$, and tensor properties such as the
data type, head dimension, and layout; it covers all attention heads,
which \system{} executes with one plan selected from statistics over all
heads.  The runtime builds the
\emph{shared mask state} $\mu(M)$ on the GPU, which holds the active-block
indices and mask statistics.  The plan selector then ranks the catalog
plans, through the request's performance regime in the plan table or
through the cost model's latency predictions, and the runtime executes the
first eligible plan (Section~\ref{sec:online-lookup}).

\section{Two-Layer Physical Plans}
\label{sec:physical-plans}

\subsection{Physical Plan}
\label{sec:plan-model}
A physical plan binds a physical mapping and a task organization to the
native CUDA implementation that realizes them.  
We represent this binding as $P=\langle g_Q,g_{KV},\sigma,e\rangle$, where $g_Q$ and $g_{KV}$ map the logical block geometry $B=(B_Q,B_{KV})$ to the physical tile geometry $T=(T_Q,T_{KV})$ on the query and key/value dimensions, respectively; the two dimensions can use different mapping ratios.
The task organization $\sigma$ assigns physical tiles to each
\emph{cooperative thread array (CTA)}, or CUDA thread block, and specifies
grouping, traversal order, the number of parallel K/V reduction
partitions, and pipeline depth.  The native CUDA implementation $e$ is one
precompiled kernel or a fixed sequence of kernels that realizes these
decisions on one GPU architecture; a common plan interface spans architectures.
\appref{C} tabulates both layers' decisions and tradeoffs.

\subsection{Physical Mapping Layer}
\label{sec:block-to-tile}
The mapping layer relates tile dimensions to logical block dimensions in
three ways: Direct, Coarsened, and Refined.

\textbf{Direct.}
Direct uses $T=B$, mapping each active logical block to one physical tile
(Figure~\ref{fig:intro-plans}(a)).
The active-block index directly identifies the tile, and all computation
within the tile corresponds to active attention entries.  Smaller logical
blocks therefore produce more, smaller tiles.

\textbf{Coarsened.}
Coarsened covers multiple logical regions with one larger physical tile
(Figure~\ref{fig:intro-plans}(b)), which reduces the tile count and reuses Q or
K/V operands across adjacent regions.  For a tile with inactive entries, \emph{membership metadata}
records the active logical blocks so that the kernel can mask inactive scores
before the softmax update; normalization and the weighted sum of values
therefore include only the entries in $E(M)$.
This reuse costs work on inactive
entries; the balance depends on the active fraction and the GPU's effective
compute and transfer rates.

\textbf{Refined.}
Refined covers a logical region with smaller physical tiles in at least
one dimension (Figure~\ref{fig:intro-plans}(c)).  Smaller tiles use fewer
registers and less shared memory in the attention-compute stage and can match the native
instruction geometry.  The costs are more tiles and, potentially, repeated
loads and state traffic.
Under a FlashAttention-style traversal, query-side refinement creates
independent query tiles, whereas K/V-side refinement updates the same query
tile's \emph{online-softmax state} sequentially across K/V tiles.  This state
holds the running maximum score, the scaled exponential sum, and the output
accumulator.  For example, one H100 Refined plan maps each $128\times128$
logical block to four $64\times64$ tiles: launch preparation expands each
active block-CSR column index into two $64$-wide K/V tile indices, and each
CTA traverses these K/V tiles for one $64$-high query tile, so the four
tiles preserve $E(M)$ (\appref{C}).

\subsection{Task Organization Layer}
\label{sec:executor-organization}

The organization layer groups and schedules physical tiles for GPU execution
according to $\sigma$; its three decisions control locality, parallelism, and
data-movement--compute overlap.

\textbf{Grouping and traversal.}
Grouping organizes active blocks into tasks by \emph{active-block runs},
maximal sequences of consecutive active K/V blocks in one query-block row.
For example, column indices $\{2,3,4,9\}$
form the runs $[2,4]$ and $[9,9]$.  Grouping combines one or more runs,
possibly from several query-block rows, into one task, and traversal
determines the order in which the CTA visits them while it keeps each query
tile and its online-softmax state on chip.
Combining more work in one task amortizes metadata accesses and increases
on-chip reuse, but leaves fewer tasks for the scheduler, and tasks with
more active blocks finish later, causing load imbalance.

\textbf{Split-K} increases parallelism by partitioning one query tile's K/V
reduction across multiple CTAs (\appref{C}).  Each CTA writes its
maximum score $m$, scaled exponential sum $l$, and partial output accumulator
$O$ to the \emph{workspace}, a temporary GPU buffer.  A subsequent kernel
rescales, merges, and normalizes these partial states, at the cost of
partial-state traffic, merge computation, and synchronization.

\textbf{Pipeline depth.}
\emph{Pipeline depth} is the number of buffered stages used to prefetch
upcoming K/V tiles; a deeper pipeline increases data-movement--compute
overlap at the cost of registers and shared memory.

\subsection{Native CUDA Catalogs}
\label{sec:catalog}

For a target architecture $A$, \system{} forms the native CUDA catalog $C_A$ by
retaining plans that have an implementation for $A$ and meet its hardware
capability requirements.  Counting each plan--logical-geometry pair as one
entry, the catalogs contain 17--22 entries across 7--10 tile shapes.

Table compilation and plan selection further filter $C_A$ per case and request
(\appref{D}).  For a request
$R=\langle A,M,B,x\rangle$, where $x$ gives the data type, head
dimension, and layout, a plan $P\in C_A$ satisfies
$\operatorname{Eligible}(P,R)$ when its implementation supports the
request's logical geometry, tensor properties, metadata format, and
available workspace while preserving the output specified by $M$.

\section{Plan Selector}
\label{sec:regime-selection}

The plan selector chooses one plan of the native CUDA catalog for each
request and moves the plan search offline.  Its two methods share
one interface.  Offline, each method turns latencies measured on the target
GPU into a way to rank plans (Section~\ref{sec:policy-construction}).
Online, the method ranks the eligible plans of a request, and a shared
ranked fallback and launch preparation turn the ranking into one prepared
plan (Section~\ref{sec:online-lookup}).  The evaluation uses the default
plan table for all speedups, and Section~\ref{sec:cost-model} compares both
methods.  \appref{G} lists the notation.

\subsection{Offline Plan-Selector Construction}
\label{sec:policy-construction}

Both methods start from the same offline profiles.  Offline profiling
records validated latencies per \emph{profiling case}, which pairs a mask
$M$ with its request parameters.  For each eligible plan $P$ in $C_A$, it
compares the output against a masked-attention reference and records the
median kernel latency as $L[M,P]$, where $M$ indexes the case.

\textbf{Plan table.}  Table compilation groups profiling cases into
performance regimes in a two-stage hierarchy.  The first stage uses the
request key $k_A(R)$ to group cases with the same architecture, logical
geometry, and tensor properties.  The second stage partitions each group
by bucketed request dimensions $u_A(R)$, such as batch size, head count,
and sequence length, and optionally by scalar mask statistics $s_A(M)$.
These statistics include \emph{active-block density}, the fraction of
logical blocks that are active, and \emph{active-block-run coverage}, the
fraction of active blocks that belong to runs of at least two, defined as
zero for an empty mask.  Each resulting partition forms a performance
regime.  The \emph{feature schema} $\mathcal{F}_A$ makes this partitioning consistent
between table compilation and runtime lookup by fixing the features, their
types, and their bucket boundaries; \appref{D} gives the compilation
algorithm.
Boundaries are predefined intervals or decision-tree thresholds (depth
chosen by source-grouped cross-validation; at least 32 cases per leaf).  The plan table $\Pi_A$ stores, for each
performance regime, the plans with valid measurements on every regime case,
ranked by geometric-mean latency.  Each ranking ends in the
key's \emph{base plan}, eligible on every profiling
case with that key.

\textbf{Calibrated cost model.}  The cost model predicts the latency of
each plan with the structure of Equation~\ref{eq:pipeline-model}.  From
$\mu(M)$ and the plan, it computes $D_P$, $F_P$, the arithmetic
$F^{\mathrm{c}}_P$ of the densest query-block row, and the tile and task
counts $N^{\mathrm{tile}}_P$ and $N^{\mathrm{task}}_P$, and predicts
\begin{equation}
  \begin{split}
  \hat{t}_P={}&\mathrm{smax}\Bigl(\frac{D_P}{\beta},\,\frac{F_P}{\pi\eta_P},\,\frac{\kappa S F^{\mathrm{c}}_P}{\pi\eta_P}\Bigr)\\
  &+\frac{c_{\mathrm{tile}}\tau_P N^{\mathrm{tile}}_P+c_{\mathrm{task}}N^{\mathrm{task}}_P}{S}+c_0,
  \end{split}
  \label{eq:cost-model}
\end{equation}
where the three terms inside $\mathrm{smax}$ are the data-movement time, the
compute time, and the time of the densest row, and the remaining terms are
per-tile and per-task overheads spread over the SMs; here $\mathrm{smax}$, a smooth maximum, is the $\ell_8$ norm, $S$ is the SM
count, and $\kappa$ weights the densest row.  Six scalars
($\beta,\pi,\kappa,c_{\mathrm{tile}},c_{\mathrm{task}},c_0$) are fitted per
GPU, and an efficiency $\eta_P$ and a tile-cost multiplier $\tau_P$ per
plan, to latencies measured on $A$.  On a new GPU, the fit starts from the
parameters of the other GPUs and uses about 100 calibration requests
(\appref{E}).

\subsection{Online Plan Selection}
\label{sec:performance-regimes}
\label{sec:online-lookup}

Online plan selection ranks the eligible plans of a request, returns one
eligible plan, and prepares it (pseudocode for the plan table in
\appref{D}).  Both methods start from the shared mask state: the runtime
builds $\mu(M)$ on the GPU, containing the \emph{block compressed sparse
row (block-CSR)} offsets and active K/V column indices of each query-block
row together with the statistics $s_A(M)$.

\textbf{Plan table: regime lookup.}  The runtime transfers only $s_A(M)$ to
the host.  The lookup function $\rho_A$ matches $k_A(R)$ in
$z_A(R)=\langle k_A(R),u_A(R),s_A(M)\rangle$, with $s_A(M)=\emptyset$ when
the plan table uses no mask statistics.  It applies the boundaries in $\mathcal{F}_A$
to identify a performance regime $r=\rho_A(z_A(R))$ and its plan ranking
$\Pi_A(r)=[P_1,\ldots,P_k]$.  Because rankings are indexed by regime,
\system{} reuses a ranking across masks in the same
regime, whereas a mask-specific cache reuses it only when the
exact mask recurs.

\textbf{Cost model: latency prediction.}  The cost model computes the
counts of each catalog plan from $\mu(M)$ and ranks the plans by
$\hat{t}_P$, so the ranking follows the structure of the current mask
rather than a profiled regime.

\textbf{Ranked fallback.} The ranking estimates performance, while
eligibility checks establish whether a plan supports the current request
and preserves its specified output.  The runtime selects the first eligible
plan in rank order; a plan-table lookup miss uses the base plan as its sole
candidate.

\textbf{Launch preparation.} The runtime derives from the shared block-CSR
only the metadata the selected plan requires---traversal descriptors,
tile membership, refinement indices, or Split-K partitions, with
workspace bindings---and invokes its native CUDA implementation in
$C_A$.

\subsection{Plan Table versus Cost Model}
\label{sec:cost-model}

The two methods cover complementary cases.  The plan table generalizes
across masks whose statistics fall in the regimes of the profiling corpus,
because its rankings are keyed on regime features rather than on mask
identity: on evaluation masks unseen during profiling, its selected plans
stay within $0.5\%$ of the
fastest plan in geometric mean (Section~\ref{sec:eval-selection}).  For
masks that differ from the corpus, a regime lookup can only
return a ranking profiled on other masks, whereas the cost model predicts
each plan's latency from the mask's counts, and its few fitted scalars
let it rank plans on a new GPU after about 100 calibration requests.

\begin{figure}[t]
  \centering
  \includegraphics[width=0.86\columnwidth]{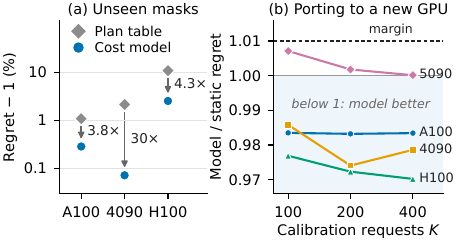}
  \caption{Calibrated cost model.  (a) Regret$-1$ of the plan table and the
    cost model on 140 unseen synthetic masks (RTX~5090 excluded: these
    masks are part of its calibration data); labels give the reduction.
    (b) Median over 20 draws of the model's regret divided by the static
    majority's, for $K$ calibration requests; dashed: 1\% non-inferiority margin.}
  \Description{Two panels.  Left: for A100, RTX 4090, and H100, arrows from the plan table's regret to the lower regret of the cost model on unseen masks, labeled 3.8x, 30x, and 4.3x.  Right: for four GPUs, the median ratio of model to static-majority regret at 100, 200, and 400 calibration requests, below the dashed non-inferiority margin.}
  \label{fig:cost-model}
\end{figure}

\textbf{Masks outside the profiled distribution.}  On 140 synthetic masks
per GPU outside the profiling distribution (seven geometries, 3--50\%
density, four row-degree and locality variants; \appref{B}), the
calibrated model has
lower regret than the plan table: $1.025$ versus $1.108$ on H100, $1.003$
versus $1.011$ on A100, and $1.001$ versus $1.021$ on RTX~4090
(Figure~\ref{fig:cost-model}(a)).

\textbf{Porting to a new GPU.}  Profiling the full catalog on one GPU takes
$3.7$--$7.6$ hours.
The cost model instead needs about 100 stratified calibration requests per
GPU: over 20 seeded draws, it meets a non-inferiority criterion against the
\emph{static majority}, each geometry's most frequent winner in the same draw, on all four GPUs (\appref{E}; Figure~\ref{fig:cost-model}(b)).
It requires a mean validation GM regret of at most 1.05 over the draws
and, in at least 12 of 20 draws, a GM regret at most 1\% above the static majority's.

\section{Implementation}
\label{sec:implementation}
\system{} consists of a Python runtime for PyTorch and a C++ and CUDA
extension.  Excluding third-party dependencies, the implementation comprises
24K lines of Python and 37K lines of C, C++, and CUDA code.
The native
catalogs cover A100, RTX~4090, H100, and RTX~5090.

\ifincludeevaluation
  \section{Evaluation}
\label{sec:evaluation}
We test whether \system{} gives high-performance plans
across masks, block geometries, and GPUs with low plan
selection and preparation overhead.
Section~\ref{sec:eval-boundary} reports request, kernel, and model speedups,
Section~\ref{sec:eval-adapt} plan variation, and
Section~\ref{sec:eval-selection} selection quality and cost.

\subsection{Experimental Setup}
\label{sec:eval-setup}

\textbf{Platforms and systems.}
On four NVIDIA GPUs (\appref{F}), each
with its catalog $C_A$ and plan table $\Pi_A$ (the default selector), we compare \system{} with
FlashInfer~\cite{ye2025flashinfer}, FlexAttention~\cite{dong2024flexattention},
and, on the four H100 families it supports,
flex-block-attn~\cite{peng2025flexblockattn}.  All runs use bfloat16 (BF16) forward attention
with head dimension 128 without CUDA graphs.  Each baseline prepares its per-mask data in every
request, and every output passes the same FP32-reference check
(\appref{F} gives the baseline configurations and tolerances).
The baselines are BSA runtimes because \system{} executes a sparsifier's
logical mask; dense kernels such as
FlashAttention-2/3~\cite{dao2024flashattention2,flashattention3} compute
every interaction, and the sparsifier chooses dense or sparse attention.

\textbf{Workloads.}
The microbenchmark contains 2,312 block-sparse video-attention masks from
four sources: Wan2.1-T2V-14B, Wan2.2-T2V-14B~\cite{wan2025wan}, and
HunyuanVideo-1~\cite{kong2024hunyuanvideo} masks generated by
SPADE~\cite{liu2026spade}, and HunyuanVideo-1.5 masks from its official
sparse-distilled image-to-video model~\cite{wu2025hunyuanvideo15}.  The
masks form nine families spanning seven logical
geometries; three use a larger query block than key/value block, reflecting
the finer key/value granularity used in sparse video-generation
systems~\cite{yang_sparsevideogen2_2025,liu2026vecattention}.
Mask \emph{density} is the fraction of active logical blocks
(Table~\ref{tab:workload-geometries}).  Q16 through Q128 abbreviate families
by query block size, and the suffixes HD and DD identify the lower- and
higher-density variants of $\mathrm{Q}_{64}\mathrm{K}_{64}$; mean densities
otherwise follow the 9--12\% default range of the four vDiT sources.  With the Q tile
kept on chip, a Direct tile performs about $4dT_QT_{KV}$ flops per
$4dT_{KV}$ bytes of BF16 K/V loads, so a family's \emph{Q-resident
intensity} is $T_Q=B_Q$ flop/byte (\appref{A}).
We also evaluate two video generation models on H100, both with
$B_Q=B_{KV}=64$: 720p Wan2.1 text-to-video with SPADE
sparsity~\cite{liu2026spade}, and 720p HunyuanVideo-1.5 image-to-video using its official sparse-distilled model~\cite{wu2025hunyuanvideo15}.

\textbf{Profiling and evaluation split.}
Each platform is profiled on about 10k requests from the same video-model
sources, and its plan table is fit on their training requests (about 8k).
A \emph{chunk} is a mask sample within a
source file; the profiling and evaluation sets do not overlap at the source-file or chunk level,
so every evaluation mask is unseen during profiling.

\begin{table}[t]
  \centering
  \caption{Workload families: mean density and masks (2,312 in total).
    $\mathrm{Q}_{q}\mathrm{K}_{k}$ has $B_Q=q$, $B_{KV}=k$, and Q-resident
    intensity $q$ flop/byte.}

  \label{tab:workload-geometries}
  \small
  \setlength{\tabcolsep}{2.5pt}
  \begin{tabular*}{\columnwidth}{@{\extracolsep{\fill}}lrrlrr@{}}
    \toprule
    Family & Density & Masks & Family & Density & Masks \\
    \midrule
    $\mathrm{Q}_{128}\mathrm{K}_{128}$  & 9.91\%  & 250 & $\mathrm{Q}_{64}\mathrm{K}_{32}$ & 10.07\% & 250 \\
    $\mathrm{Q}_{128}\mathrm{K}_{64}$   & 9.91\%  & 250 & $\mathrm{Q}_{32}\mathrm{K}_{32}$ & 9.76\%  & 250 \\
    $\mathrm{Q}_{64}\mathrm{K}_{64}$-HD & 5.35\%  & 250 & $\mathrm{Q}_{32}\mathrm{K}_{16}$ & 9.89\%  & 250 \\
    $\mathrm{Q}_{64}\mathrm{K}_{64}$    & 11.95\% & 312 & $\mathrm{Q}_{16}\mathrm{K}_{16}$ & 9.79\%  & 250 \\
    $\mathrm{Q}_{64}\mathrm{K}_{64}$-DD & 18.20\% & 250 &                                  &         &     \\
    \bottomrule
  \end{tabular*}
\end{table}

\textbf{Latency definitions.}
\emph{Complete-request latency} includes each backend's
preprocessing and execution through completion, and
\emph{kernel execution latency} covers the selected native CUDA
implementation.
Model results report the total latency of all BSA requests and of a 50-step diffusion loop.

\textbf{Metrics.}
Speedup is baseline latency divided by \system{} latency, and
\emph{family speedup} is the ratio of mean latencies over paired cases.  The
\emph{geometric mean} (GM) of the nine family speedups is the
nine-family GM, and that of per-case speedups is the paired-case GM.
\emph{Regret} is the kernel latency of the selected plan over that of the
fastest measured eligible plan in $C_A$ for the same case (1$\times$ is optimal); P95 and P99
denote the 95th and 99th percentiles, and the tie band is 3\%.  The 95\%
bootstrap confidence intervals over masks of all nine-family GMs are within
$\pm5\%$ and exclude 1 (\appref{F}).

\subsection{Speedups for Requests, Kernels, and Models}
\label{sec:eval-boundary}

\textbf{Complete requests.}  \system{} reduces nine-family GM complete-request latency for every
platform--baseline pair (Figure~\ref{fig:request-speedup-gm}).
Nine-family GM
complete-request speedups are $1.57$--$1.76\times$ over FlashInfer and
$2.01$--$3.16\times$ over FlexAttention across the four GPUs.
The GM speedup over flex-block-attn is $2.00\times$ on its four
H100 families.

\begin{figure}[t]
  \centering
  \includegraphics[width=0.92\columnwidth]{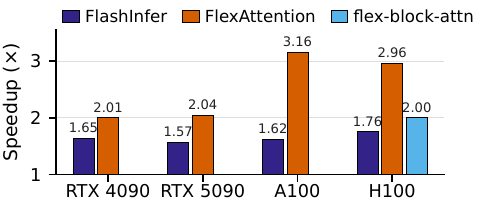}
  \caption{Nine-family GM complete-request speedup (baseline/\system{}
    latency) on four GPUs; flex-block-attn: four H100 families.
    Dashed line: 1$\times$.}
  \Description{Grouped bars for RTX 4090, RTX 5090, A100, and H100 show the nine-family geometric-mean complete-request speedup of \system{} over FlashInfer (1.57 to 1.76), FlexAttention (2.01 to 3.16), and, on H100, flex-block-attn (2.00).}
  \label{fig:request-speedup-gm}
\end{figure}

\textbf{Kernel execution.}  \system{} also reduces kernel execution latency across the measured
platform--baseline pairs (Table~\ref{tab:kernel-speedup}): nine-family GM
kernel speedup is $1.16$--$1.59\times$ over FlashInfer and
$2.24$--$4.91\times$ over FlexAttention; over flex-block-attn, which
accepts only $B=(64,64)$ and $(128,128)$, the GM speedup on these four H100
families is $3.50\times$.  Over the fastest Direct plan of each mask, the
selected plans are $1.04$--$1.17\times$ faster on every GPU.
Kernel latency excludes all preparation, so it measures execution
alone, as when a baseline reuses its plan or block mask.

\begin{table}[t]
  \centering
  \caption{Kernel speedup (baseline/\system{} latency), nine-family GM.
    H100 flex-block-attn covers four families; ``--'': unmeasured pair.}
  \label{tab:kernel-speedup}
  \small
  \begin{tabular*}{\columnwidth}{@{\extracolsep{\fill}}lrrrr@{}}
    \toprule
    Baseline        & RTX~4090      & RTX~5090      & A100          & H100          \\
    \midrule
    FlashInfer      & 1.16$\times$ & 1.16$\times$ & 1.19$\times$ & 1.59$\times$ \\
    FlexAttention   & 2.56$\times$ & 2.24$\times$ & 4.25$\times$ & 4.91$\times$ \\
    flex-block-attn & --            & --            & --            & 3.50$\times$ \\
    \bottomrule
  \end{tabular*}
\end{table}

\textbf{Model workloads.}  \system{} accelerates both attention and the full diffusion loop in the two
H100 model workloads (Figure~\ref{fig:model-workload}).  For Wan2.1 and
Hunyuan\-Video-1.5, speedups over FlashInfer and flex-block-attn are
$2.17$--$2.88\times$ for BSA requests, $1.63$--$3.13\times$ for
kernel execution, and $1.21$--$1.24\times$ for the 50-step diffusion
loop.
Loop gains are smaller because non-attention
computation is unchanged: with block-sparse attention, attention takes
about 32\% of the Wan2.1 loop under FlashInfer and 18\% under \system{},
so the attention time saved alone accounts for a loop speedup of about
$1.21\times$, close to the measured $1.23\times$.

\begin{figure}[t]
  \centering
  \includegraphics[width=0.85\columnwidth,trim=0 3bp 0 2bp,clip]{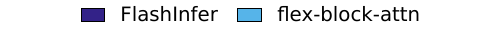}\par
  \begin{subfigure}[t]{0.340\columnwidth}
    \centering
    \includegraphics[width=\linewidth,trim=0 17bp 0 2bp,clip]{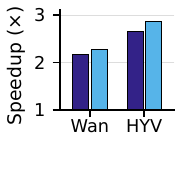}\par
    \caption{BSA request.}
    \label{fig:model-workload-attention}
  \end{subfigure}\hfill
  \begin{subfigure}[t]{0.311\columnwidth}
    \centering
    \includegraphics[width=\linewidth,trim=0 17bp 0 2bp,clip]{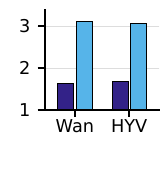}\par
    \caption{Kernel only.}
    \label{fig:model-workload-kernel}
  \end{subfigure}\hfill
  \begin{subfigure}[t]{0.311\columnwidth}
    \centering
    \includegraphics[width=\linewidth,trim=0 17bp 0 2bp,clip]{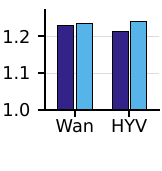}\par
    \caption{Diffusion loop.}
    \label{fig:model-workload-diffusion}
  \end{subfigure}\par
  \caption{Model speedups on H100 (baseline/\system{} latency)
    over FlashInfer and flex-block-attn
    for Wan2.1 (Wan) and HunyuanVideo-1.5 (HYV); (c) covers the 50-step
    diffusion loop.}
  \Description{Three side-by-side subfigures use grouped vertical bars to show speedup for Wan2.1 and HunyuanVideo-1.5 against FlashInfer and flex-block-attn, with one shared legend and one shared y-axis title. Panel a reports BSA request speedup, panel b reports attention kernel-execution speedup, and panel c reports 50-step diffusion-loop speedup. Each panel marks one times speedup with a dashed horizontal line.}
  \label{fig:model-workload}
\end{figure}

\subsection{Plan Performance Varies with Geometry, Mask, GPU, and Organization}
\label{sec:eval-adapt}

\textbf{Mapping types.}  Direct, Coarsened, and Refined each yield the fastest measured plan in part
of the workload (Figure~\ref{fig:plan-dependence}).
Without Coarsened plans, even the fastest remaining plan per mask
has kernel GM regret $1.119$ on A100 and $1.105$ on H100.

\begin{figure*}[t]
  \centering
  \includegraphics[width=0.86\linewidth]{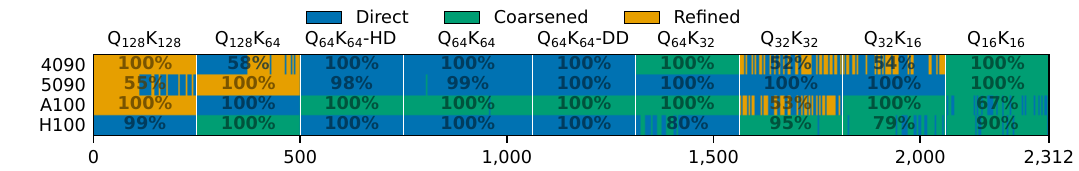}\par
  \caption{Mapping type of the fastest measured plan for 2,312 masks
    on four GPUs.  Families are ordered by Direct-tile Q-resident intensity,
    from 128 flop/byte for Q128 to 16 for Q16 (Table~\ref{tab:workload-geometries}).
    Percentages: share of the most frequent mapping type per region.}
  \Description{A four-row strip plot shows the fastest mapping type for the same masks on RTX 4090, RTX 5090, A100, and H100. Blue, green, and orange denote Direct, Coarsened, and Refined. Workload-family names appear above the plot, and each platform--family region is annotated with its majority-mapping share.}
  \label{fig:plan-dependence}
\end{figure*}

\textbf{Logical geometry.}  On H100, geometry changes the fastest mapping: low-intensity
Q16 and Q32 families favor Coarsened, Q64 families mostly Direct,
and the Q128 families split between Direct ($\mathrm{Q}_{128}\mathrm{K}_{128}$)
and Coarsened ($\mathrm{Q}_{128}\mathrm{K}_{64}$).

\textbf{Mask.}  Mask variation changes the fastest mapping even at fixed geometry and GPU:
within the H100 $\mathrm{Q}_{32}\mathrm{K}_{16}$ family,
197 masks favor Coarsened and 53 favor Direct
(Figure~\ref{fig:plan-dependence}).

\textbf{Density.}  In the three $\mathrm{Q}_{64}\mathrm{K}_{64}$ families, the fastest
mapping follows the GPU across densities: Coarsened wins 100\%, 99.7\%,
and 100\% of A100 cases beyond the 3\% tie band (HD, default, DD), and
Direct wins on H100.

\textbf{GPU architecture.}  GPU architecture changes the fastest mapping type for the same masks.  The
$\mathrm{Q}_{32}\mathrm{K}_{16}$ masks favor Coarsened on A100, whereas 135 of
the same masks favor Direct on RTX~4090 (\appref{B}), and RTX~5090 favors
Direct for most of them (Figure~\ref{fig:plan-dependence}).

\textbf{Task organization.}  Task organization adds speedup at a fixed mapping: with the H100
$\mathrm{Q}_{128}\mathrm{K}_{128}$ masks at one Refined mapping, the
organization that \system{} selects yields a $1.10\times$ paired-case GM kernel speedup
over the base organization.  It stays within $1.0\%$ of the fastest
measured organization per case and improves 84 of 250 masks beyond the
3\% tie band, slowing none (\appref{F}).

\subsection{Plan Selection Preserves Gains at Low Cost}
\label{sec:eval-selection}
\textbf{Selection quality.}  Plans selected by \system{} have GM regret at or below 1.005 with the
plan table and 1.004 with the cost model on every platform
(Figure~\ref{fig:selector-regret}), that is, at most 0.5\% higher kernel latency
in geometric-mean terms, even as the fastest mapping varies across masks
and GPUs (Figure~\ref{fig:plan-dependence}).  With the plan table, selected
plans are within 3\% of the minimum measured eligible-plan latency in the
native CUDA catalog on 94.3--99.7\% of the 2,312 masks per platform, and
P99 regret ranges from 1.007 to 1.103; with the cost model, the
corresponding values are 96.1--99.7\% and 1.016--1.096.
A fixed plan per GPU and logical geometry (the most frequent training
winner) has P99 regret 1.46--1.67 on A100, RTX~4090, and H100 and slows
143--286 masks by over 10\%, versus 1.02--1.10 and 1--28 for the plan
table; on RTX~5090 the two are close (P99 1.03 versus 1.01).  Even on the
H100 $\mathrm{Q}_{64}\mathrm{K}_{64}$ families, where Direct always wins,
the fixed plan has GM regret 1.04--1.09 versus at most 1.007.

\begin{figure}[t]
  \centering
  \includegraphics[width=0.85\linewidth]{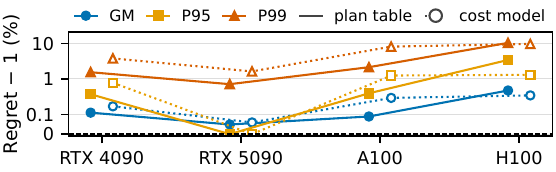}\par
  \caption{Regret$-1$ (symlog scale; 0 is optimal): GM, P95, and P99 over
    2,312 masks.  Solid: plan table; dotted: cost model.}
  \Description{A single-column line plot reports geometric-mean, P95, and P99 selected-plan regret minus one, in percent on a logarithmic axis, for RTX 4090, RTX 5090, A100, and H100, for the plan table (solid) and the cost model (dotted, hollow markers), with a dashed reference line at zero.}
  \label{fig:selector-regret}
\end{figure}

\textbf{Selection overhead.}  Every request pays for choosing and preparing
its execution.  We measure each system's complete preparation, everything
before its kernel launch (mask processing, selection or planning, and
launch preparation), in a separate synchronized pass of each
$\mathrm{Q}_{64}\mathrm{K}_{64}$ request.  Across the four GPUs, its median
is 1.2--5.4~ms for \system{} and 2.4--12.3~ms for FlashInfer, whose
\texttt{plan()} runs for every new mask and head, so the preparation of
\system{} is $1.95$--$2.35\times$ faster.
Within it, plan selection takes $82.5$~$\mu$s on average on H100
(\appref{F} splits the rest).

\begin{table}[tbp]
  \centering
  \caption{Complete-request speedup between plan-selection variants on 270 H100 evaluation masks, nine-family GM.}
  \label{tab:request-cost-ablation}
  \small
  \setlength{\tabcolsep}{3pt}
  \begin{tabular}{lrrr}
    \toprule
    Population & $t_{\mathrm{A0}}/t_{\mathrm{A1}}$ & $t_{\mathrm{A1}}/t_{\mathrm{A2}}$ & $t_{\mathrm{A0}}/t_{\mathrm{A2}}$ \\
    \midrule
    Cold unique  & $0.083\times$ & $14.42\times$ & $1.192\times$ \\
    Warm repeats & $1.160\times$ & $2.003\times$ & $2.323\times$ \\
    Amortized    & $0.268\times$ & $7.009\times$ & $1.877\times$ \\
    \bottomrule
  \end{tabular}
\end{table}

\textbf{Search versus selection.}  The A0/A1/A2 ablation measures how selection quality and cost combine in
complete requests on a fixed sample of 270 H100 evaluation masks
(Table~\ref{tab:request-cost-ablation}).
A0 uses Direct mapping with $T=B$, the fixed per-geometry plan of
specialized kernels.  A1 uses
\emph{exhaustive online search}, preparing and probing every eligible native
CUDA catalog candidate (median of five timed runs each) before executing
the first candidate within 1\% of the fastest.
A2 uses \system{} with the
same native CUDA catalog and the fixed H100 plan table, preparing only the
selected plan.  Plans selected by A2 match those selected by A1 in kernel execution
speedup over A0 across the nine families (\appref{F}).
A \emph{cold unique request} is the first occurrence of a
request, including selection, launch preparation, any probes,
and execution; \emph{warm exact repeats} are passes 2--4 of the
same request; amortized latency averages passes 1--4.

\textbf{Cold and warm requests.}  \system{} serves cold unique requests $14.42\times$ faster than A1
and $1.192\times$ faster than Direct
(Table~\ref{tab:request-cost-ablation}), whereas A1 reaches only
$0.083\times$ over Direct because it prepares and probes every eligible
plan for each cold request.  \system{} instead reuses plan-table rankings
across different masks in the same performance regime
(Section~\ref{sec:online-lookup}).
On warm exact repeats, neither method probes candidates, but A1 still
prepares its cached winner while A2 reuses the prepared plan state, so
\system{} is $2.003\times$ faster; over four passes it is $7.009\times$
faster than A1.

\fi
\section{Related Work}
\label{sec:related}

To our knowledge, \system{} is the first specialized runtime system
for kernel-level physical planning of dynamic block-sparse attention
in video generation.
Coarsening, sub-tiling, and split-K are established kernel techniques, and
existing systems can express some of these mappings: FlexAttention, for
example, can coarsen through a block mask with a larger block size and a
token-level mask function.  \system{} differs in three ways: it keeps the logical mask
fixed and provides Direct, Coarsened, and Refined as alternative
physical plans for it (Section~\ref{sec:block-to-tile}); it selects the mapping and organization
for each request from request features and mask statistics, so a ranking
serves masks that were never profiled (Section~\ref{sec:regime-selection});
and it builds a native catalog per GPU, because the fastest implementation
of the same mapping differs across GPUs (Section~\ref{sec:why-ranking-changes}).

\textbf{Sparse attention for video diffusion.}
Video sparsifiers, including SPADE with input-adaptive blocking, select
active interactions from spatial--temporal patterns, semantic similarity,
or learned
sparsity~\cite{xi_sparse_2025,yang_sparsevideogen2_2025,zhang_vsa_2025,longcat2025technical,wu2025hunyuanvideo15,tan2025dsv,liu2026spade};
\system{} takes their logical masks and block geometries.

\textbf{Sparse compilers and input-aware selection.}
Sparse tensor compilers generate code from composable formats and
schedules, including format
decomposition~\cite{ye2023sparsetir,won2025insum,kjolstad2017taco};
sparse GPU kernels reorder rows to balance SpMM and SDDMM
work~\cite{gale2020sputnik}; and per-matrix autotuning and learned format or
kernel selectors choose an implementation for each sparse
input~\cite{vuduc2005oski,zhao2018bridging,won2023waco,swann2024seer}.
These techniques target additive reductions, and most amortize each choice
over runs on the same input; Seer instead predicts a kernel from runtime
input features.  BSA uses an online softmax, and no mask recurs in the
measured generations (Section~\ref{sec:intro}); \system{} therefore precompiles semantics-preserving
mappings and keys rankings on mask statistics, not identity.

\textbf{Autotuning and learned cost models.}
Tensor compilers and kernel languages autotune schedule and tile
parameters, often with a learned cost
model~\cite{tillet2019triton,chen2018autotvm,zheng2020ansor,zhai2023tlp},
and amortize the search over many runs of one operator configuration;
\system{} amortizes profiling over performance regimes keyed by request
features and mask statistics, and its cost model fits two parameters per
plan to the pipeline terms.

\textbf{Attention runtimes.}
Block Sparse Attention and flex-block-attn provide kernels for preset
block sizes~\cite{guo2024blocksparse,peng2025flexblockattn};
FlashInfer plans and schedules at runtime over multiple tile
configurations~\cite{ye2025flashinfer}, and FlexAttention compiles
programmable mask variants with tiles that match or subdivide logical
blocks in its standard Triton forward path~\cite{dong2024flexattention}.
AttentionEngine and FlashMask extend attention kernels to diverse masks
and hardware, and db-SP and FVAttn balance attention work across
GPUs~\cite{chen_attentionengine_2025,wang_flashmask_2025,chen2025dbsp,liu2026fvattn}.
These runtimes tie physical tiles to logical blocks or recompute a
mask-specific plan for every new mask, and PSA splits and merges logical blocks into execution tiles for
multi-level pooled KV representations~\cite{li2025psa}; \system{} instead
selects among mappings that retain, combine, or subdivide logical blocks
for each request from a native catalog per GPU.

\section{Conclusion}
\label{sec:conclusion}

\system{} adapts BSA execution to the mask and GPU through physical mapping and task organization, and chooses plans with low overhead via a profiled plan table or a calibrated cost model.
On four GPU generations and 2,312 real attention masks, it reduces request latency against every baseline; in two industrial video models, it speeds up BSA requests by up to $2.88\times$ and the 50-step loop by $1.21$--$1.24\times$.
Its plan table selects plans within $0.5\%$ of the fastest catalog plan in geometric mean, and its calibrated cost model ports plan selection to a new GPU with about 100 calibration requests.

\nocite{*}
\bibliographystyle{ACM-Reference-Format}
\bibliography{references}

\appendix

\section{Data--Compute Pipeline Model}
\label{app:pipeline}

\begin{figure}[t]
  \centering
  \includegraphics[width=.9\columnwidth]{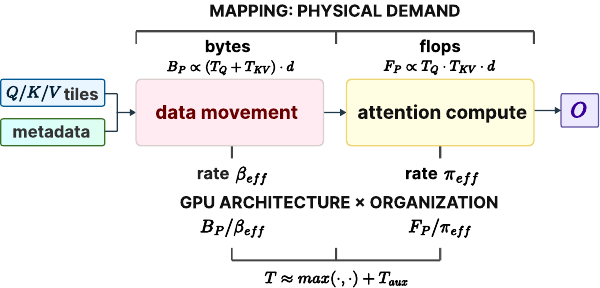}
  \caption{Block-sparse attention as a two-stage data--compute pipeline.  The
    plan and GPU jointly determine byte demand $D_P(M)$, arithmetic demand
    $F_P(M)$, their effective service rates, and auxiliary cost.}
  \Description{A two-stage producer--consumer pipeline.  Q/K/V tiles and
    metadata enter a data-movement stage, and an attention-compute stage
    produces the output O.  Each stage has a demand and an effective rate;
    the estimated execution time is the larger of the two stage times plus
    an auxiliary cost.}
  \label{fig:pipeline}
\end{figure}

Figure~\ref{fig:pipeline} illustrates the two-stage data--compute pipeline
behind Equation~\ref{eq:pipeline-model}.

On-chip reuse amortizes traffic that occurs once per query-tile traversal.
A FlashAttention-style traversal keeps one Q tile, its online-softmax state,
and its output accumulator on chip while successive K/V tiles pass through
the mainloop.  Under simplified bfloat16 (BF16) accounting that includes
K/V loads but excludes metadata and cache effects, each physical tile
performs about $4dT_QT_{KV}$ floating-point operations across $QK^\top$ and
the multiplication of attention weights by $V$, while loading about
$4dT_{KV}$ K/V bytes.  The Q load and final-output store occur once per
traversal.  Under this simplified accounting, their per-tile contribution
diminishes as more K/V tiles are visited, and the ratio of matrix arithmetic
to traffic approaches $T_Q$ floating-point operations per byte.

Physical mapping and task organization jointly affect demand and the balance
between the data-movement and attention-compute stages.  Larger tiles can
reduce tile and indexing counts but add arithmetic when they cover inactive
entries.  Smaller tiles can reduce such
inactive work and per-task footprint but increase tile count and repeated
operand traffic.  Task grouping, traversal order, parallel decomposition,
and pipeline depth further change locality, reuse, parallelism, overlap, and
auxiliary cost.  Their combined effects determine the physical cost of
executing the same $E(M)$.

\section{Mapping Choices Across GPUs and Masks}
\label{app:sankey}

This section supports Sections~\ref{sec:why-ranking-changes},
\ref{sec:cost-model}, and~\ref{sec:eval-adapt} of the main paper:
a same-GPU implementation comparison, the fastest mapping of the same masks
across GPUs, and the fastest mapping on synthetic masks.

\subsection{Same GPU and Mapping, Different Implementation}

Table~\ref{tab:impl-control} fixes the GPU (H100), the logical geometry
$B=(128,128)$, and 60 masks, and compares the Refined plan with two Direct
kernels.  The 4-warp \texttt{mma.sync} Direct kernel written for SM8x loses
to Refined on every mask.  A warp-specialized SM90 Direct kernel uses
\texttt{wgmma}~\cite{nvidia_cutlass_wgmma} and \texttt{setmaxnreg}, an SM90
instruction that moves registers from the data-movement warps to the
compute warps~\cite{flashattention3} so that they can hold the accumulators
of a large tile; it outperforms Refined on every mask except one tie.  In a
separate run, it beats Refined on 44 of 56 non-tied masks with
\texttt{setmaxnreg} and on none of 59 without it.

\begin{table}[t]
  \centering
  \caption{Same GPU and mapping, different implementation: H100,
    $B=(128,128)$, 60 masks; wins within 3\% are ties.  The last row
    counts non-tied masks of a separate run: 56 with and 59 without
    \texttt{setmaxnreg}.}
  \label{tab:impl-control}
  \small
  \setlength{\tabcolsep}{2pt}
  \begin{tabular*}{\columnwidth}{@{\extracolsep{\fill}}lr@{}}
    \toprule
    Comparison & Wins \\
    \midrule
    SM8x mma.sync Direct vs.\ Refined: Refined wins & 60/60 \\
    SM90 wgmma Direct vs.\ Refined: Direct wins & 59/60 \\
    Separate run, w/ $\rightarrow$ w/o setmaxnreg & 44/56 $\rightarrow$ 0/59 \\
    \bottomrule
  \end{tabular*}
\end{table}

\subsection{Fastest Mapping Across GPUs}

\begin{figure}[t]
  \centering
  \includegraphics[width=0.8\linewidth,trim=0 5bp 0 4bp,clip]{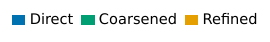}\par
  \includegraphics[width=0.8\linewidth,trim=0 4bp 0 3bp,clip]{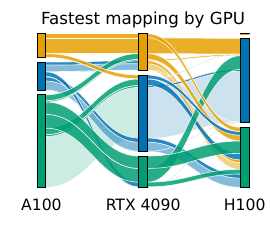}
  \caption{The fastest physical mapping for 2,312
    masks depends on GPU architecture.  Ribbon opacity increases
    with the number of GPUs that support multiple measured mapping types.}
  \Description{A three-stage Sankey showing how the fastest
    mapping changes for the same masks across A100, RTX 4090, and H100; paler
    ribbons indicate fewer measured alternatives.}
  \label{fig:intro-sankey}
\end{figure}

Figure~\ref{fig:intro-sankey} traces how the mapping type of the fastest
measured plan changes for the same masks across A100, RTX~4090, and
H100 (Section~\ref{sec:eval-adapt} of the main paper).

\subsection{Performance Regimes on Synthetic Masks}

\begin{figure*}[t]
  \centering
  \includegraphics[width=\textwidth]{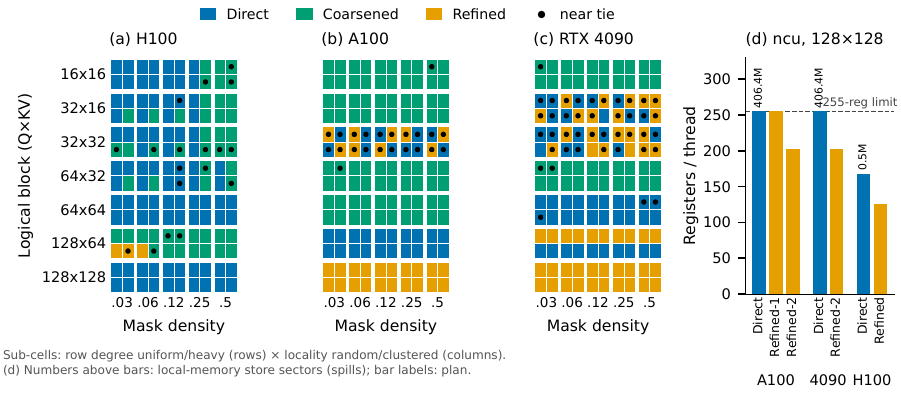}
  \caption{Performance regimes on synthetic masks.  (a)--(c) Mapping type
    of the fastest plan in each GPU's native CUDA catalog for 140 synthetic masks
    (7 logical geometries $\times$ 5 densities $\times$ 4 row-degree and
    locality variants); a dot marks a near tie (best and runner-up within the
    repeat spread).  Timings are kernel-boundary medians of 3 repeats.  The
    best mapping type changes with geometry, mask, and GPU.  (d) Nsight Compute
    counters at logical $128\times128$ (one balanced mask, $d=0.25$):
    registers per thread and spill (local-store) traffic per plan.  Panel (d)
    reports counters, not profiled time.}
  \Description{Three panels show, for A100, RTX 4090, and H100, the mapping
    type of the fastest plan for each synthetic mask by logical geometry;
    a fourth panel shows registers per thread and local-store sectors of the
    Direct and Refined kernels at 128 by 128.}
  \label{fig:regime-a0}
\end{figure*}

Figure~\ref{fig:regime-a0} shows the mapping type of the fastest plan on
140 synthetic masks per GPU, together with the register and spill counters
of the Direct and Refined kernels at $128\times128$.
These masks also form the out-of-distribution set of
Section~\ref{sec:cost-model} of the main paper.

\section{Physical Plans and Native CUDA Catalogs}
\label{app:plans}

\subsection{Plan Decisions}
Table~\ref{tab:plan-hierarchy} summarizes the decisions of the mapping and
organization layers (Sections~\ref{sec:block-to-tile}
and~\ref{sec:executor-organization} of the main paper), their
execution outputs, and their main tradeoffs.

\begin{table}[tbp]
  \caption{Physical-plan decisions and performance tradeoffs.}
  \label{tab:plan-hierarchy}
  \centering
  \small
  \setlength{\tabcolsep}{4pt}
  \begin{tabularx}{\columnwidth}{@{}>{\hsize=.70\hsize\linewidth=\hsize}Y
    >{\hsize=.95\hsize\linewidth=\hsize}Y
    >{\hsize=1.35\hsize\linewidth=\hsize}Y@{}}
      \toprule
\textbf{Decision} &
\multicolumn{1}{c}{\textbf{Output}} & \multicolumn{1}{c@{}}{\textbf{Tradeoff}} \\
      \midrule
      \multicolumn{3}{@{}l}{\textit{Physical mapping layer~(main \S\ref{sec:block-to-tile})}} \\
      \addlinespace[2pt]

      \textbf{Direct}
      & One tile per active logical block
      & Simple indexing vs.\ more tiles for small blocks \\
      \addlinespace[2pt]

      \textbf{Coarsened}
      & Larger tiles with membership metadata
      & Operand reuse vs.\ compute on inactive entries \\
      \addlinespace[2pt]

      \textbf{Refined}
      & Smaller tiles along refined dimensions
      & Resource savings vs.\ more tiles and repeated traffic \\

      \midrule
      \multicolumn{3}{@{}l}{\textit{Task organization layer~(main \S\ref{sec:executor-organization})}} \\
      \addlinespace[2pt]

      \textbf{Grouping and} \textbf{traversal}
      & Grouped tasks and tile traversal order
      & Locality and metadata reuse vs. coarse scheduling and imbalance \\
      \addlinespace[2pt]

      \textbf{Split-K}
      & Parallel K/V-reduction tasks and a merge
      & Parallelism vs.\ partial-state traffic, merging, and synchronization \\
      \addlinespace[2pt]

      \textbf{Pipeline} \textbf{depth}
      & Buffered tile prefetch stages
      & Load--compute overlap vs. register and shared-memory use \\

      \bottomrule
  \end{tabularx}
\end{table}

\subsection{A Refined Plan}
One H100 plan maps each $128\times128$ logical block to four
$64\times64$ physical tiles.  The block-CSR indices store the active blocks
of each query-block row.
Launch preparation expands each block-CSR column index, which names a
$128$-wide active block, into the indices of two $64$-wide K/V tiles.  The
kernel launch assigns each CTA one $64$-high query tile, so the two query
halves of a logical block run in separate CTAs, and each CTA processes its
query tile while traversing the expanded K/V tiles.  Together, the four tiles
cover the original logical block and preserve $E(M)$.

\subsection{Split-K}
\begin{figure}[t]
  \centering
  \includegraphics[width=.8\columnwidth]{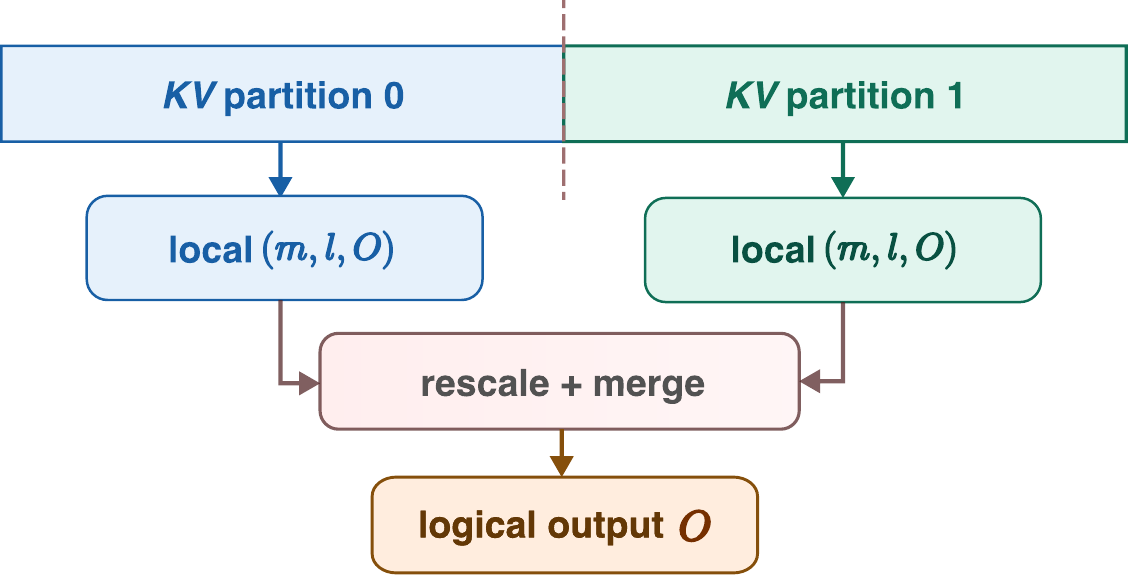}
  \caption{Split-K organization.  CTAs process partitions of one query
    tile's K/V reduction domain and write local $(m,l,O)$ states to the
    workspace.  A reduction kernel rescales and merges these states to
    produce the output.}
  \Description{The K/V tiles for one query tile are divided among parallel
    CTA tasks.  Each task writes a local maximum, scaled exponential sum,
    and output accumulator to the GPU workspace.  A reduction kernel
    combines the partial states into the final output.}
  \label{fig:splitk}
\end{figure}

Figure~\ref{fig:splitk} shows the Split-K organization described in
Section~\ref{sec:executor-organization} of the main paper.

\subsection{Native CUDA Catalogs}
The native CUDA library that holds a catalog $C_A$ takes 36--58~MB: 36~MB on
RTX~5090, 44~MB on RTX~4090, 47~MB on A100, and 58~MB on H100.

\section{Plan-Table Compilation and Online Selection}
\label{app:compile}

\subsection{Table Compilation}

Algorithm~\ref{alg:compile} gives offline table compilation.  Lines~2--9
record validated latencies $L[M,P]$; lines~10--13 form the request-key
groups and the feature-schema partitions; lines~14--19 rank the plans of
each performance regime and append the base plan.

For the second stage of the hierarchy, the
boundaries of the feature schema $\mathcal{F}_A$ are predefined intervals or
thresholds learned by a decision tree, whose leaves define the finer
partitions.  We select the tree depth through cross-validation on corpus
folds.  Masks from the same source remain in one fold, and each leaf
contains at least 32 cases.

\begin{algorithm}[t]
  \caption{Table compilation for target architecture $A$.}
  \label{alg:compile}
  \small
  \begin{algorithmic}[1]
    \Require native CUDA catalog $C_A$, profiling corpus $\mathcal M$,
    feature schema $\mathcal{F}_A$
    \Ensure plan table $\Pi_A$
    \State $L\gets\emptyset$; $\Pi_A\gets\emptyset$
    \Statex \textit{Profiling}
    \ForAll{cases $R=\langle A,M,B,x\rangle$ in $\mathcal M$}
    \ForAll{$P\in C_A$ with $\operatorname{Eligible}(P,R)$}
    \State Run $P$ and compare with the reference output
    \If{the output passes numerical validation}
    \State $L[M,P]\gets$ median kernel latency
    \EndIf
    \EndFor
    \EndFor
    \Statex \textit{Regimes and rankings}
    \ForAll{request keys $k$ in $\mathcal M$}
    \State $\mathcal M_k\gets$ cases with $k_A(R)=k$
    \State $P_{\mathrm{base}}\gets$ catalog's base plan for $k$
    \State $\mathcal B_k\gets$ partition $\mathcal M_k$ using $\mathcal{F}_A$
    \ForAll{nonempty buckets $b\in\mathcal B_k$}
    \State $\mathcal C_b\gets$ plans with a valid $L[M,P]$
    \Statex \hspace{\algorithmicindent}\hspace{\algorithmicindent}%
    \hspace{\algorithmicindent}for every case $M$ in bucket $b$
    \If{$\mathcal C_b\ne\emptyset$}
    \State Sort $\mathcal C_b$ by increasing geometric-mean latency over $b$
    \State Break exact ties by plan identifier
    \State $\Pi_A[k,b]\gets\mathcal C_b$ followed by $P_{\mathrm{base}}$
    \EndIf
    \EndFor
    \EndFor
    \State \Return $\Pi_A$
  \end{algorithmic}
\end{algorithm}

\subsection{Online Plan Selection}
Algorithm~\ref{alg:select} gives the online selection steps of
Section~\ref{sec:online-lookup} of the main paper.  Lines~1--5 build
$\mu(M)$ and transfer only the required scalar statistics $s_A(M)$ to the
host; lines~6--9 form $z_A(R)$ and look up the regime; lines~10--12 handle a
lookup miss with the base plan; and lines~13--18 perform ranked fallback.

\begin{algorithm}[t]
  \caption{Online plan selection for $R=\langle A,M,B,x\rangle$.}
  \label{alg:select}
  \small
  \begin{algorithmic}[1]
    \Require plan table $\Pi_A$, native CUDA catalog $C_A$,
    feature schema $\mathcal{F}_A$
    \Ensure selected plan $P_{\mathrm{sel}}$ or \texttt{no-eligible-plan}
    \State Build GPU $\mu(M)$: block-CSR and $\mathcal{F}_A$ summaries
    \State $s_A(M)\gets\emptyset$
    \If{$\mathcal{F}_A$ requires summaries}
    \State Copy required scalars from $\mu(M)$ to host $s_A(M)$
    \EndIf
    \State $z\gets\langle k_A(R),u_A(R),s_A(M)\rangle$
    \State $r\gets\rho_A(z)$
    \If{lookup finds $r$}
    \State $\mathcal L\gets\Pi_A(r)$
    \Else
    \State $\mathcal L\gets[\text{base plan for }k_A(R)]$
    \EndIf
    \ForAll{$P\in\mathcal L$ in rank order}
    \If{$\operatorname{Eligible}(P,R)$}
    \State \Return $P$
    \EndIf
    \EndFor
    \State \Return \texttt{no-eligible-plan}
  \end{algorithmic}
\end{algorithm}

\section{Cost-Model Protocol}
\label{app:cost-model}

This section records the protocol behind
Section~\ref{sec:cost-model} of the main paper.

\subsection{Model Form and Fitting}
The model predicts the median kernel
latency of a plan with the form of Section~\ref{sec:policy-construction}.  Its
six per-GPU scalars are $\beta$, $\pi$, $\kappa$, $c_{\mathrm{tile}}$,
$c_{\mathrm{task}}$, and $c_0$; each plan adds $\eta_P$ and $\tau_P$.  The fit
minimizes a soft-$\ell_1$ loss (scale 0.1) of $\log(\hat{t}_P/t_P)$ over the
measured latencies $t_P$, plus a quadratic penalty (weight 0.3) that pulls
the logarithms of $\eta_P$ and $\tau_P$ toward a prior; a full fit uses the
prior $\eta_P=\tau_P=1$.  Cross-validation on the training requests of the
profiling corpus selects this form on every GPU.  Regret is measured on
the validation requests of the profiling corpus.  A full fit has 38--52
parameters per GPU and uses 16,146--23,600 measured latencies of the
training requests, about 390--510 measurements per parameter; a
100-request calibration draw contains about 225--300 measured latencies,
and the prior from the other GPUs regularizes its per-plan parameters.

\subsection{Calibration Quality}
Per-plan calibration lowers the GM regret
(Section~\ref{sec:eval-setup} of the main paper) relative to
uncalibrated estimates on the profiling corpus's validation requests: a
peak-rate roofline estimate reaches $1.084$--$1.219$, an analytical model
with the pipeline terms but no per-plan calibration reaches
$1.066$--$1.093$, and the calibrated model reaches $1.001$--$1.003$, within
$0.2\%$ of the plan table ($1.000$--$1.003$) on every GPU.

\subsection{Calibration-Set Sampling}
On a new GPU, the fit starts from the
other three GPUs.  The compute and overhead scalars are rescaled from their
fits by the ratios of peak throughput and clock frequency, and each
per-plan parameter starts at the median of the values fitted for the same
plan on those GPUs; the per-plan penalty pulls toward these values.  A
calibration set of $K$ requests is drawn from the training requests by
stratified sampling: $K$ is split across logical geometries in proportion
to their training-request counts (largest remainder, at least one request
per geometry when $K$ allows), and requests are drawn uniformly within each
geometry.  The fit uses only the calibration requests.  The static majority
fitted on the same draw ranks, for each logical geometry, the plans by the
number of calibration requests on which each plan is fastest.

\subsection{Non-Inferiority Criterion}
The calibration experiment compares the
calibrated model with the static majority fitted on the same draw, over 20
seeded draws of the calibration set.  On each GPU, the criterion passes
when the mean over the 20 draws of the model's validation regret GM is at
most 1.05 and, in at least 12 of the 20 draws, the model's GM is at most
the static majority's GM times $1+\delta$, with $\delta=0.01$ (about twice
the 0.5\% timing-noise floor of the profiling data).  Calibration with
100 requests drawn by stratified sampling passes this criterion on all
four GPUs.

\section{Evaluation Details}
\label{app:eval}

\begin{table}[t]
  \centering
  \caption{Hardware platforms: per-SM peak dense bfloat16 (BF16) Tensor
    Core throughput with FP32 accumulation (BF16/SM, TFLOP/s), per-SM peak
    global-memory bandwidth (BW/SM, GB/s), their ratio (F/B, flop/byte),
    and rated maximum board power (Cap, W).}
  \label{tab:hardware-platforms}
  \small
  \setlength{\tabcolsep}{2pt}
  \begin{tabular*}{\columnwidth}{@{\extracolsep{\fill}}llrrrr@{}}
    \toprule
    GPU & Memory & BF16/SM & BW/SM & F/B & Cap \\
    \midrule
    RTX~4090 (SM89)  & 24GB & 1.291 &  7.88 & 163.9 & 450 \\
    RTX~5090 (SM120) & 32GB & 1.232 & 10.54 & 116.9 & 575 \\
    A100 PCIe (SM80) & 80GB & 2.889 & 17.92 & 161.2 & 300 \\
    H100 SXM (SM90a) & 80GB & 7.496 & 25.38 & 295.4 & 700 \\
    \bottomrule
  \end{tabular*}
\end{table}

\begin{table*}[t]
  \centering
  \caption{Nine-family GM speedup (baseline/\system{} latency) with 95\%
    bootstrap confidence intervals over masks.}
  \label{tab:bootstrap-ci}
  \small
  \begin{tabular}{@{}lcccc@{}}
    \toprule
    Baseline & RTX~4090 & RTX~5090 & A100 & H100 \\
    \midrule
    \multicolumn{5}{@{}l}{\textit{Complete request}} \\
    FlashInfer & 1.647 [1.619, 1.670] & 1.571 [1.549, 1.590] & 1.622 [1.597, 1.643] & 1.758 [1.710, 1.798] \\
    FlexAttention & 2.007 [1.974, 2.035] & 2.045 [2.026, 2.061] & 3.160 [3.106, 3.206] & 2.964 [2.886, 3.032] \\
    flex-block-attn & -- & -- & -- & 2.004 [1.926, 2.078] \\
    \midrule
    \multicolumn{5}{@{}l}{\textit{Kernel}} \\
    FlashInfer & 1.161 [1.155, 1.166] & 1.157 [1.153, 1.162] & 1.188 [1.183, 1.194] & 1.590 [1.580, 1.599] \\
    FlexAttention & 2.565 [2.534, 2.593] & 2.243 [2.226, 2.258] & 4.248 [4.194, 4.300] & 4.907 [4.847, 4.963] \\
    flex-block-attn & -- & -- & -- & 3.502 [3.463, 3.536] \\
    \bottomrule
  \end{tabular}
\end{table*}

\subsection{Hardware Platforms}
Table~\ref{tab:hardware-platforms} lists the four GPUs of
Section~\ref{sec:eval-setup} of the main paper.  The H100 host has two
Intel Xeon Platinum 8481C CPUs (2.7~GHz), the A100 and RTX~4090 share a host
with two Intel Xeon Gold 6230R CPUs (2.1~GHz), and the RTX~5090 host has an
AMD Ryzen~7 9800X3D (up to 5.27~GHz).  Each run is pinned to a fixed set of
host cores (12 cores of one NUMA node on H100), and every speedup compares
systems on the same host.
A100 and RTX~4090 favor different mappings despite nearly equal peak
compute-to-bandwidth ratios (161 and 164 flop/byte), so this ratio alone
does not determine the fastest mapping (Section~\ref{sec:eval-adapt} of
the main paper).

\subsection{Baseline Configurations and Correctness Check}
FlexAttention builds its block mask from the block-sparse layout
(\texttt{BlockMask.from\_kv\_blocks}, $64\times64$ blocks) for every request
and runs a kernel compiled with \texttt{torch.compile} and warmed up before
timing.  For each GPU, geometry, and metric, FlashInfer~0.6.6 uses the
wrapper that is fastest on profiling masks among its eligible wrappers.
Kernel latency uses the paged or the variable block-sparse wrapper.  The
request path uses the block-sparse wrapper (the variable one for
$B_Q=128$ on RTX~5090), which shares one mask across heads and runs one
\texttt{plan()} per head; the variable wrapper, which plans all heads at
once, has higher request latency on the first eight profiling masks of
every H100 geometry it supports ($11.9$ versus $10.4$~ms for
$B=(128,128)$, $15.4$ versus $14.6$ for $(128,64)$, $28.8$ versus $16.9$
for $(64,32)$, and $13.6$ versus $8.2$ for $(64,64)$) and does not support
some masks of the three Q16 and Q32 geometries; the paged wrapper is slower
still on every geometry.
flex-block-attn~0.1 (commit bc2ec1ee) builds its
coarse mask during request preparation, so its kernel latency covers only
its forward kernel.  On H100 and RTX~4090, all systems run on
PyTorch~2.9.1 with CUDA~12.8, and on RTX~5090 on PyTorch~2.13.0 with
CUDA~13.1.  No system uses CUDA graphs.
Every output of every system passes the same check against an FP32 masked
reference: its maximum error is at most twice that of a BF16 reference plus
$10^{-5}$, and its elementwise error is at most $0.05+0.05|o|$ for reference
value $o$.

\subsection{Confidence Intervals}
Table~\ref{tab:bootstrap-ci} gives 95\% bootstrap confidence intervals for
every nine-family GM speedup of Section~\ref{sec:eval-boundary} (four
families for flex-block-attn).  Each of 5,000 resamples draws masks with
replacement within each family and repeats the headline aggregation.

\subsection{Task-Organization Ablation}
To isolate the organization layer, we hold the H100
$\mathrm{Q}_{128}\mathrm{K}_{128}$ mapping to an exact $2\times2$ cover of
physical $64\times64$ tiles across all 250 masks.
O0 uses the fixed base task organization; O1 uses the task organization
selected by \system{}; O2 uses the \emph{oracle}, the fastest measured eligible
task organization for each case.  Task organization yields a $1.099\times$
paired-case GM kernel execution speedup over O0 at the same mapping: O1
stays within $1.0\%$ of O2 and improves 84 cases over O0 by more than the
3\% tie band (Table~\ref{tab:organization-ablation}).

\begin{table}[tbp]
  \centering
  \caption{Task-organization ablation on H100 at fixed mapping
    ($\mathrm{Q}_{128}\mathrm{K}_{128}$, 250 masks).  Kernel
    speedup is paired-case GM over O0; regret is relative to O2.  CI denotes
    confidence interval; W/T/L counts wins/ties/losses against O0.}
  \label{tab:organization-ablation}
  \small
  \begin{tabular}{@{}lccc@{}}
    \toprule
    Variant        & Speedup [95\% CI]    & Regret & W/T/L    \\
    \midrule
    O0: fixed base & 1.000                & 1.1095 & --       \\
    O1: \system{}   & 1.099 [1.078, 1.121] & 1.0100 & 84/166/0 \\
    O2: oracle     & 1.109 [1.090, 1.131] & 1.0000 & --       \\
    \bottomrule
  \end{tabular}
\end{table}

\subsection{Request-Path Breakdown}
A separate, synchronized preparation pass of each request splits the
complete preparation of \system{}
(Section~\ref{sec:eval-selection} of the main paper).  On the
$\mathrm{Q}_{64}\mathrm{K}_{64}$ requests, its median total is 3.2~ms on
H100, 5.2~ms on A100, 5.4~ms on RTX~4090, and 1.2~ms on RTX~5090, close to
the timed complete-request minus kernel latency of \system{}.  On H100, building the
shared mask state $\mu(M)$ on the GPU takes 0.91~ms, deriving the mask
statistics $s_A(M)$ 0.05~ms, plan selection 0.08~ms, deriving the row
lists of the selected plan 0.40~ms, and building its launch structures and
workspace 1.34~ms; synchronization and Python code between these steps
take the remaining 0.39~ms.

\subsection{Plan Quality of Online Search and \system{}}
\begin{figure}[t]
  \centering
  \includegraphics[width=0.95\linewidth]{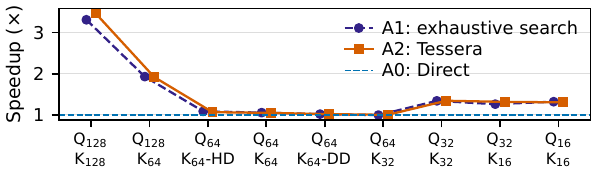}\par
  \caption{Kernel execution speedup over Direct (A0) for
    A1 and \system{} (A2), by workload family on 270 H100
    evaluation masks (269 paired).  Speedup is A0 latency divided by
    variant latency.}
  \Description{A single-column line plot shows A1 and A2 kernel execution speedups over Direct for nine workload families.}
  \label{fig:plan-quality-ablation}
\end{figure}

Figure~\ref{fig:plan-quality-ablation} breaks down by workload family the
kernel execution speedups of A1 (exhaustive online search) and A2
(\system{}) over A0 (Direct) from Section~\ref{sec:eval-selection} of
the main paper.  The nine-family GM is $1.373\times$ for A2 and
$1.365\times$ for A1, a difference within A1's 1\% band.

\section{Notation}
\label{app:notation}

Table~\ref{tab:notation} lists the notation of the main paper.

\begin{table}[t]
  \centering
  \caption{Notation of the main paper.}
  \label{tab:notation}
  \small
  \setlength{\tabcolsep}{3pt}
  \begin{tabularx}{\columnwidth}{@{}l>{\raggedright\arraybackslash}X@{}}
    \toprule
    Symbol & Meaning \\
    \midrule
    $M$, $E(M)$ & Logical mask; its effective interaction set \\
    $B=(B_Q,B_{KV})$ & Logical block geometry \\
    $T=(T_Q,T_{KV})$ & Physical tile geometry \\
    $A$ & Target architecture (GPU) \\
    $R=\langle A,M,B,x\rangle$ & Runtime request; $x$: tensor properties \\
    $P=\langle g_Q,g_{KV},\sigma,e\rangle$ & Physical plan: mappings $g_Q,g_{KV}$, task organization $\sigma$, native CUDA implementation $e$ \\
    $t_A(P,M)$ & Pipeline-model latency of $P$ on $A$ \\
    $D_P$, $F_P$ & Bytes and arithmetic of plan $P$ \\
    $\beta^{\mathrm{eff}}_A$, $\pi^{\mathrm{eff}}_A$, $t_{\mathrm{aux}}$ & Effective bandwidth and compute rate; auxiliary time \\
    $C_A$ & Native CUDA catalog of $A$ \\
    $\operatorname{Eligible}(P,R)$ & $P$ supports $R$ and preserves its output \\
    $L[M,P]$ & Profiled median kernel latency \\
    $k_A(R)$, $u_A(R)$ & Request key; bucketed request dimensions \\
    $s_A(M)$ & Scalar mask statistics \\
    $\mathcal{F}_A$ & Feature schema (features and bucket boundaries) \\
    $z_A(R)$, $\rho_A$ & Lookup features; lookup function \\
    $r$, $\Pi_A$, $\Pi_A(r)$ & Performance regime; plan table; its ranking \\
    $\mu(M)$ & Shared mask state (block-CSR and statistics) \\
    $\hat{t}_P$ & Latency predicted by the cost model \\
    $\eta_P$, $\tau_P$ & Per-plan efficiency and tile-cost multiplier \\
    \bottomrule
  \end{tabularx}
\end{table}

\clearpage

\end{document}